\PassOptionsToPackage{hyphens}{url}
\documentclass[final,3p,times]{elsarticle}
\usepackage{graphicx}
\usepackage{hyperref}
\usepackage{breakurl}
\usepackage{etoolbox}
\usepackage{booktabs}
\usepackage{amssymb}
\usepackage{float}
\usepackage{eurosym}
\usepackage{amsmath}
\usepackage{xcolor}

\hypersetup{
    hidelinks,
    breaklinks=true
}

\begin{document}

\begin{frontmatter}

\title{Spatial equity and decentralization trade-offs in deep decarbonization of the European power system}

\author{Kristoffer Hedegaard Aden\corref{cor1}, Alexander Kies}
\cortext[cor1]{Corresponding author}
\affiliation{organization={Department of Electrical and Computer Engineering, Aarhus University},
             city={Aarhus},
             country={Denmark}}

\begin{abstract}
Standard EU energy system modelling approaches optimize for least-cost,
leading to highly centralized systems, in conflict with political feasibility and
physical security concerns.
This paper incorporates decentralisation as a constraint in a European energy system model using a novel, linear load-weighted renewable capacity constraint, the K-parameter, which scales with total system renewable capacity to avoid interference with decarbonisation targets.
The model is a 37-node electricity-only brownfield system based on the
PyPSA-EUR framework, with projected 2050 loads and technology costs.
A total of 105 optimized scenarios are analyzed at 14 levels of decarbonization
and 8 levels of decentralization.
Full decarbonization leads to an 80\% cost increase due to, among other factors, a 78\% increase in energy generation capacity.
Without decentralisation constraints, system equity initially improves but collapses at high decarbonisation levels due to concentration in regions with optimal renewable resources.
Moderate decentralization of K=7 achieves 76\% of the equity benefits at only a 9\% cost increase compared to K=1.
This indicates that moderate decentralization can be a viable strategy to balance societal preferences and cost-efficiency in the European energy transition.
\end{abstract}


\end{frontmatter}

\section{Introduction}

\noindent
In a world where the necessity of reducing the carbon emissions of our energy
networks is no longer a suggestion, but a requirement, careful and deliberate
planning of the energy system of the future is needed.
When the Paris Climate Agreement was signed into action in 2015, it signaled the commitment
of the United Nations to reducing their overall carbon footprint \cite{paris2015}.
Achieving these targets presents a major technological and infrastructural challenge.

\medskip
\noindent
Somewhat unique to the issue of decarbonization in the European Union is the
fact that, although the Paris Climate Agreement signifies the common goal of
decarbonization, the EU is still comprised of many different countries with many
different political ideologies, laws, societal preferences, and goals of their own.
This means that, although it has long been known that a shared and interconnected
energy grid is the most efficient and economically preferable solution \cite{rodriguez2015},
reality becomes a lot more complicated as the member nations also must have their
own best interests in mind.

\medskip
\noindent
Historically, the impact of the union being made up of independent national actors
has often been discarded, when trying to construct the electricity grid of the
future, if for no other reason than the fact that implementing social preferences
and societal paradigms in strict mathematical models of the European energy system
is a challenging problem \cite{pfenninger2014}.
These models, which typically minimise total system cost under techno-economic assumptions, are often used to evaluate future energy infrastructure pathways and policy strategies. While such approaches have provided important insights into European decarbonisation planning \cite{henke2022}, recent work has increasingly highlighted that socio-political factors also play a critical role in shaping feasible and publicly acceptable transition pathways \cite{dioha2023beyond}. Nevertheless, unconstrained least-cost optimisation frequently leads to highly centralised energy systems, where renewable deployment becomes concentrated in regions with the most favourable resources \cite{eriksen2017}.
Previous studies investigating regional equity, decentralisation, and energy autonomy in highly renewable European systems have similarly demonstrated that increasing spatial self-sufficiency generally introduces additional infrastructure and system cost trade-offs \cite{neumann2021costs,kendziorski2022centralized}. Although centralised systems are often economically efficient, they may increase infrastructure vulnerability, cross-border dependence, and broader energy security concerns \cite{cherp2011three}.

\medskip
\noindent
While centralization as a strategy is often the most cost-efficient, it leads to a
suite of other challenges.
Two such challenges are those of physical security and political feasibility.
With the recent surge in so-called hybrid attacks on European energy infrastructure
\cite{iiss2025}, the focus on physical security has increased.
If a system is overly centralized, it means that a large amount of critical
infrastructure is installed in a relatively small area.
This makes the infrastructure an obvious target for such hybrid attacks, as a single
hostile action can impact more infrastructure and more people.
As for political feasibility, nations who end up as net energy importers become less
self-sufficient and increasingly reliant on the net exporting regions.
This shift in power balance heavily favors the net exporters and, as such, may face
resistance from net importing nations that wish to remain at a high degree of
self-sufficiency.
Recent literature has increasingly highlighted the importance of integrating equity and justice-related considerations into energy system modelling frameworks, including quantitative decision-support approaches for energy justice and fair allocation principles in energy system design \cite{heleno2022optimizing,vaagero2023can,goforth2025incorporating}. More broadly, recent work has emphasized that a just energy transition requires explicit representation of societal and distributional objectives within quantitative assessment frameworks \cite{rios2025just}. At the same time, previous work has highlighted the methodological difficulty of incorporating such societal objectives into large-scale techno-economic optimisation models \cite{lonergan2023energy}.

Recent work has further demonstrated that different fairness allocation principles can substantially influence optimal energy system configurations and infrastructure deployment patterns \cite{vaagero2024effects}. In particular, spatial distributive justice considerations have increasingly been identified as important factors in renewable energy infrastructure siting and regional energy planning \cite{lehmann2024spatial}. A key challenge is therefore introducing decentralization constraints without fundamentally interfering with decarbonization targets or compromising computational tractability in large-scale scenario exploration.

This paper introduces a linear load-weighted renewable capacity constraint, referred to as the K-constraint, which enables systematic exploration of decentralization within a European power system optimization framework while preserving linearity and scalability. The methodology is implemented in a 37-node European electricity system model based on the PyPSA-Eur framework and evaluated across 105 scenarios spanning multiple levels of decarbonization and decentralization.
The main research questions are:

\begin{itemize}
    \item[(I)]   How does enforcing decentralization alter least-cost decarbonization pathways in a European electricity system?
    \item[(II)]  Which system mechanisms drive the interaction between decarbonization and decentralization?
    \item[(III)] What trade-offs emerge between spatial equity, system cost, and infrastructure deployment when decentralization is enforced?
\end{itemize}
The results reveal a pronounced transition between approximately 80–90\% decarbonization, where the system shifts from a load-following planning regime toward a resource-quality-driven regime characterized by increasing spatial concentration of renewable deployment, storage expansion, and transmission dependence. Furthermore, the analysis demonstrates that moderate decentralization can achieve substantial improvements in spatial equity at relatively limited additional system cost.

The remainder of the paper is structured as follows. Section 2 presents the modelling framework and the proposed K-constraint methodology. Section 3 presents the scenario results and analyzes the interactions between decarbonization and decentralization. Section 4 discusses the implications, limitations, and broader relevance of the findings, while Section 5 concludes the paper.
\section{Methodology}

\subsection{Modelling framework}

\begin{figure}[htbp]
    \centering
    \includegraphics[width=0.85\textwidth]{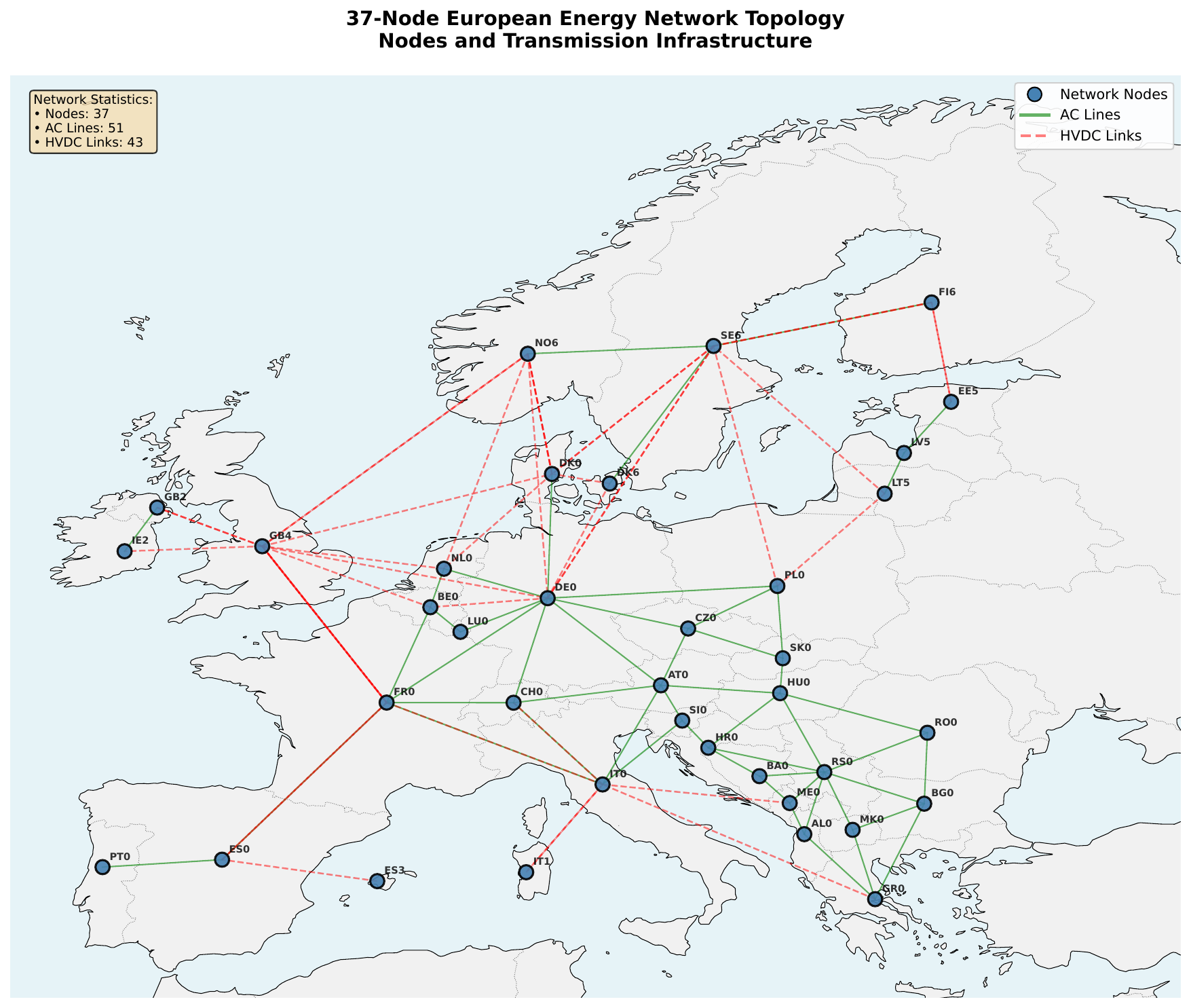}
    \caption{Overview of the 37-node European energy network, showing nodes and interconnecting AC transmission lines and HVDC links.}
    \label{fig:network}
\end{figure}

\noindent
The analysis is performed using the open-source framework PyPSA-Eur \cite{hoersch2023}, which formulates the European electricity system as a linear optimization problem. This model has previously been applied to large-scale studies of highly renewable European energy systems, including sector-coupled optimisation and transmission expansion analyses \cite{brown2018synergies}.The study uses a clustered 37-node representation of the European transmission system with hourly temporal resolution and projected 2050 electricity demand. The reduced spatial resolution was chosen to enable large-scale scenario exploration while preserving continental-scale transmission structure.
Studies of similar power system models show relatively small sensitivity to
increases and decreases in the number of nodes \cite{schyska2021}, and it was found
that a time-scale of 1-hour resolution was sufficient to capture the majority of
time-dependent effects in the system.
The resulting network topology is shown in Figure~\ref{fig:network}.

\subsection{Load rescaling and cost assumptions}

\noindent
The precompiled network models come preloaded with electrical demand time series
for each of the nodes from the ENTSOE-E transparency platform \cite{entsoe}.
These hourly demand time series are, however, in the case of the precompiled model,
from the year 2013.
Each node is assigned a target electricity consumption value based on figures from
CarbonFree Europe \cite{carbonfree}.
These are then used together with the load values from 2013 to calculate a nodal
scaling factor for each node:

\begin{equation}
    \text{scaling factor}_n = \frac{\text{target Load}_n}{\text{2013 Load}_n} \qquad \forall\, n
    \label{eq:scaling}
\end{equation}

\noindent
These nodal scaling factors are then used as weights on the hourly load time series
for each node, to produce a rescaled load time series for each node:

\begin{equation}
    \text{rescaled Load}_{n,t} = \text{scaling factor}_n \cdot \text{2013 Load}_{n,t} \qquad \forall\, n,\, t
    \label{eq:rescaledload}
\end{equation}

\noindent
This way the total load is scaled to the projected value but the general shape of
the load time series is preserved.
This is of course a simplification as continuing electrification of different
sectors, such as heating and transport, as well as changing weather patterns due
to climate change, may very well alter the consumption pattern in the future.
However, sector coupling and synthetic climate change models are beyond the scope of
this paper and this simplification is therefore necessary.
Technology cost assumptions are based on the Danish Energy Agency technology catalogues \cite{dea2023} and IEA data sources \cite{iea2023}, with detailed techno-economic parameters provided in Appendix~A.

\subsection{Optimisation strategy and scenario design}
In order to be able to analyze the impacts of varying decarbonization constraints,
and decentralization measures, a baseline system that is subjected to neither
decarbonization nor decentralization constraints must first be optimized.
The objective function for finding the cheapest possible electrical system is defined
as minimizing the sum of all costs:

\begin{equation}
    \min \sum_{n} \text{Capital costs} + \sum_{n,t} \text{Variable costs}
    \label{eq:objective}
\end{equation}

\noindent
In order to construct a feasible solution space the objective function must be
constrained by a set of bounding constraints.
Initially an equality constraint is defined so that, for each snapshot, all nodes
must have their electricity demand met by a combination of energy production and
nodal balancing:

\begin{equation}
    \text{generation}_{n,t} + \text{balance}_{n,t} = \text{demand}_{n,t} \quad \leftrightarrow \quad \lambda_{n,t} \qquad \forall\, n,\, t
    \label{eq:loadbalance}
\end{equation}

\noindent
where $\lambda$ is known as the dual-variable or the Karush-Kuhn-Tucker (KKT)
multiplier associated with the constraint.
This variable can be interpreted as the shadow price, or marginal price of
electricity in node $n$ at time $t$.
Once the baseline system has been optimised, decarbonised systems are defined by adding a CO\textsubscript{2} emissions constraint:

\begin{equation}
    \text{total emissions}_C \leq (1 - C) \cdot \text{baseline emissions} \quad \leftrightarrow \quad \mu_C
    \label{eq:co2}
\end{equation}

\noindent
Where $C$ is the level of decarbonization as a fraction of the baseline emissions
and $\mu$ is the KKT multiplier and can be interpreted as the price of emitting an
additional unit of CO\textsubscript{2} in the system at decarbonization level $C$.
The following levels of $C$ were chosen in order to achieve a wide range of
decarbonization, while increasing resolution at higher levels of decarbonization,
where the system tends to change more rapidly:

\begin{equation}
    C \in \{0\%, 10\%, 20\%, 30\%, 40\%, 50\%, 60\%, 70\%, 80\%, 90\%, 93\%, 96\%, 99\%, 100\%\}
    \label{eq:cvalues}
\end{equation}

\subsection{The decentralisation constraint --- the K-parameter}

\noindent
The novelty of this work lies in the systematic exploration of interactions between
decarbonization and decentralization using a load weighted renewable capacity
constraint (K-parameter). A key feature of the formulation is that the decentralization constraint scales with total renewable deployment in the system. This avoids the decentralization constraint implicitly tightening the decarbonization target at higher renewable penetrations and allows consistent comparison across decarbonization levels while preserving linearity and computational tractability.
As discussed above, decentralization is necessary in order to ensure a spatially balanced and resilient European energy system.
The decentralization of the systems was constrained using a novel approach based on
\cite{eriksen2017} as seen in:

\begin{equation}
    \frac{1}{K} \cdot \omega_n \cdot R_{\text{total}} \leq R_{\text{cap},n} \leq K \cdot \omega_n \cdot R_{\text{total}} \qquad \forall\, n
    \label{eq:kconstraint}
\end{equation}

\noindent
Where $\omega_n = L_n / \sum_{n'} L_{n'}$ is the normalized load-share weight for
node $n$, $R_{\text{total}}$ is the total renewable capacity across all nodes,
$R_{\text{cap},n}$ is the installed renewable capacity at node $n$, and $K$ is the
decentralization parameter. In the limiting case $K \rightarrow \infty$, the renewable deployment is unconstrained and follows the least-cost solution. As K approaches 1, renewable deployment becomes increasingly proportional to nodal load share, corresponding to progressively stronger spatial equity enforcement.
This way each node is given a minimum renewable capacity as a fraction of the total
renewable capacity in the system based on their share of the total load.
Simultaneously each node is also assigned a maximum renewable capacity as a factor
of the total renewable capacity in the system based on their share of the total
load.
By then gradually having $K$ approach 1 the system is forced to decentralize until
at $K = 1$ each node is required to install renewable capacity exactly equal to
their load-share weighted renewable capacity, which leads to the system becoming
increasingly decentralized.

\medskip
\noindent
This approach has the benefit of being linear, but also avoids another pitfall when
implementing decentralization in systems that also are subjected to decarbonization
constraints.
It is for this reason that the decentralization constraint is defined as a fraction
of the total renewable capacity in the system, ensuring that the decentralization
constraint scales with the amount of renewable capacity needed to meet the
decarbonization constraint.
Based on preliminary experiments the following levels of $K$ were chosen:

\begin{equation}
    K \in \{\infty,\, 7,\, 6,\, 5,\, 4,\, 3,\, 2,\, 1\}
    \label{eq:kvalues}
\end{equation}

\noindent
The resulting scenario ensemble therefore consists of 105 optimized networks, each at a different level of decarbonization and decentralization.
To measure decentralisation outcomes, the Standard Deviation of Nodal Renewable Capacity per Load (SDRCL) is used as the primary equity metric:

\begin{equation}
    \text{SDRCL} = \sqrt{\frac{1}{n} \sum_{n} \left( \frac{R_{\text{cap},n}}{L_n} - \mu \right)^2} \qquad \text{where} \quad \mu = \frac{1}{n} \sum_{n} \frac{R_{\text{cap},n}}{L_n}
    \label{eq:sdrcl}
\end{equation}

\noindent
A low value of SDRCL indicates that the renewable capacity is more evenly
distributed across the nodes relative to their load, while a high value indicates
a more uneven distribution.
SDRCL is used here as a proxy for spatial equity, where low values correspond to renewable deployment approximately proportional to nodal load, while high values indicate increasing spatial concentration of renewable capacity. The objective of the scenario ensemble is not to provide a definitive European planning blueprint, but rather to identify structural interactions between deep decarbonization and spatial equity constraints in large-scale electricity system optimization.

\section{Results}

\subsection{Baseline system}

\noindent
As a reminder the baseline system is purely constrained by the objective function
(\ref{eq:objective}) and the load balance constraint (\ref{eq:loadbalance}).
The mismatch between the load and the fossil generation has been filled with
renewable generation, as the fossil based generation technologies are not allowed
to expand their capacity.
The unconstrained baseline system already exhibits substantial spatial concentration of renewable deployment, with renewable capacity primarily installed in regions with favourable resource conditions. Although fossil-based generation remains present, renewable electricity supplies slightly more than half of total generation. Balancing is primarily achieved through transmission expansion and battery storage deployment, while hydrogen storage remains economically unattractive under the assumed technology costs.

\begin{figure}[htbp]
    \centering
    \includegraphics[width=0.85\textwidth]{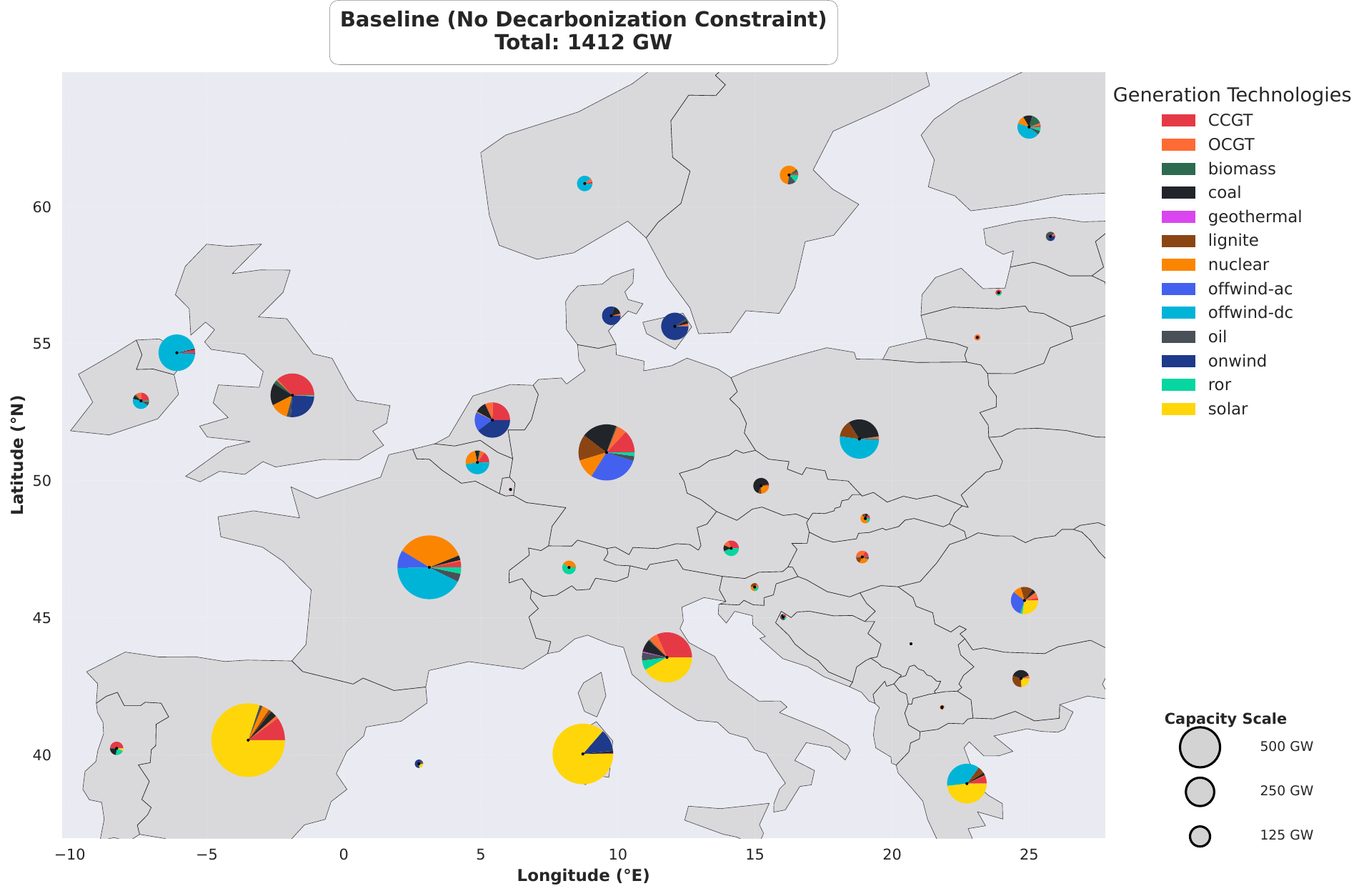}
    \caption{Baseline generator capacity map, showing installed generation capacity by technology and node.}
    \label{fig:baseline_map}
\end{figure}

\begin{figure}[htbp]
    \centering
    \includegraphics[width=0.85\textwidth]{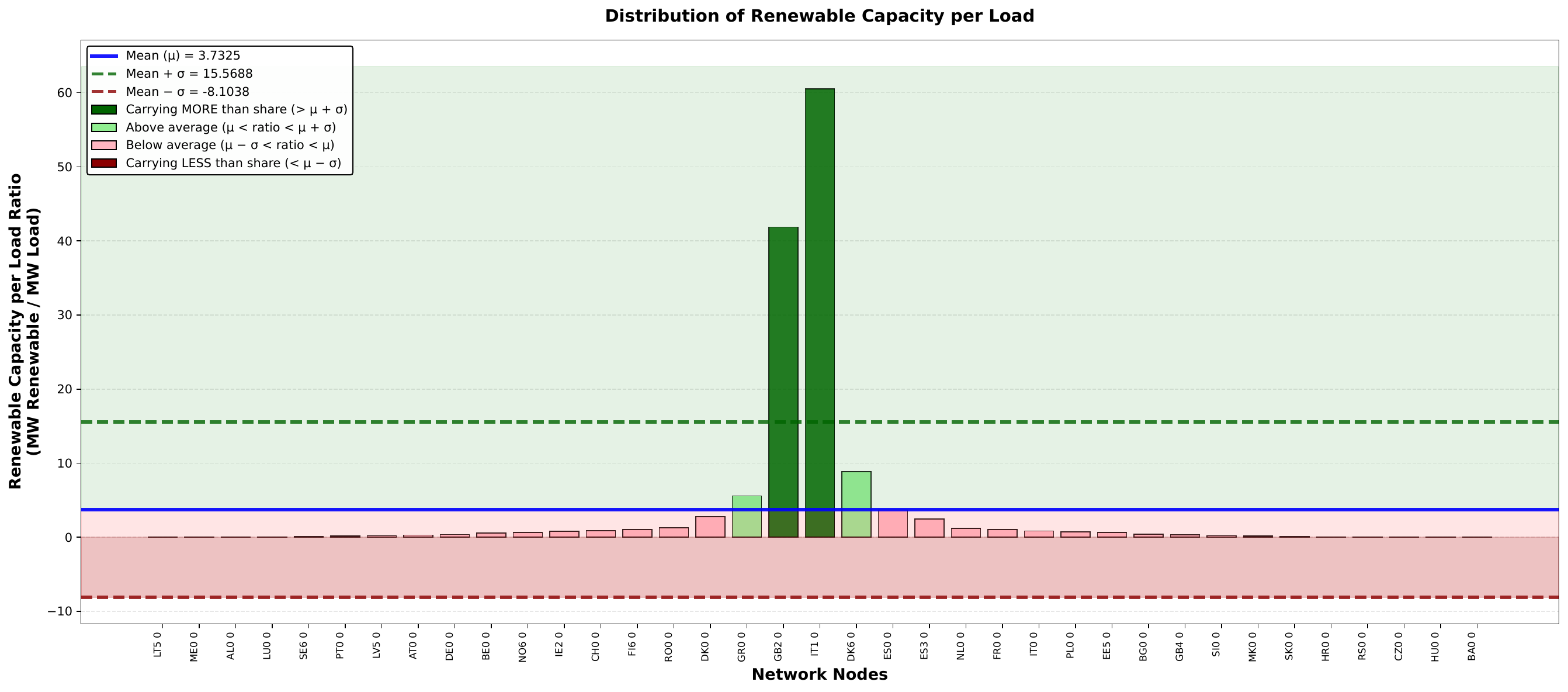}
    \caption{Baseline renewable capacity per load distribution with standard deviation bounds, illustrating the high degree of centralisation in the unconstrained system. SDRCL~=~11.84.}
    \label{fig:baseline_sdrcl}
\end{figure}

\medskip
\noindent
The mean marginal demand-weighted price of electricity in the baseline system is
calculated to be 68.56~\euro/MWh which is fairly reasonable and well within the
same order of magnitude as found in papers on similar models \cite{eriksen2017}.
The baseline system exhibits a SDRCL value of 11.84 MW$_{\text{renewable}}$/MW$_{\text{load}}$, indicating substantial spatial concentration of renewable deployment relative to nodal load distribution (Figures~\ref{fig:baseline_map} and~\ref{fig:baseline_sdrcl}).
While the heavily centralized nature of the system is optimal from a simple
least-cost perspective, it will likely lead to issues with public acceptance,
political feasibility and security.

\subsection{Decarbonised systems}

\begin{figure}[htbp]
    \centering
    \includegraphics[width=0.85\textwidth]{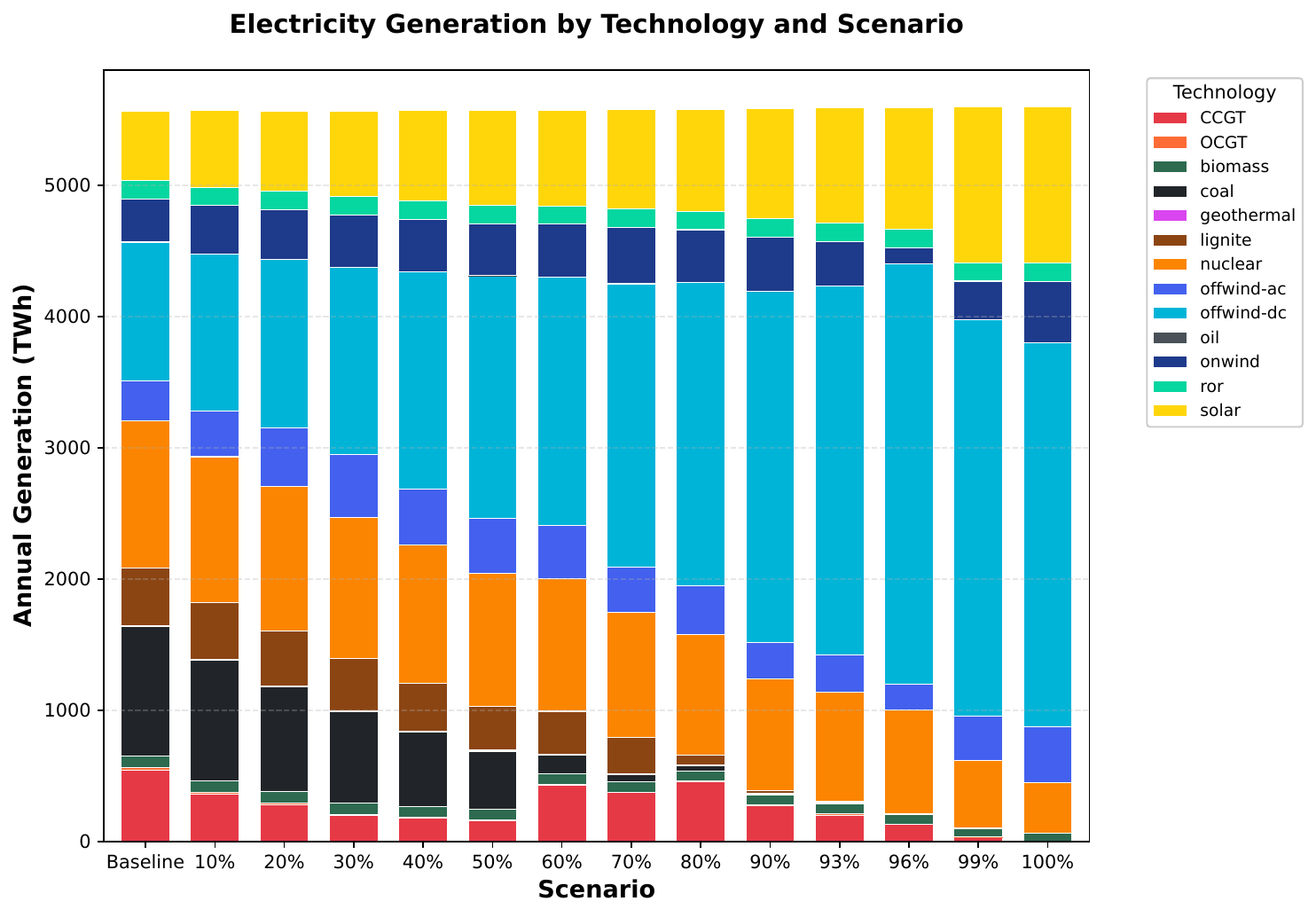}
    \caption{Technology generation trends across decarbonisation scenarios, showing the dominance of offshore wind at high decarbonisation levels and the decline of nuclear generation above 80\% CO\textsubscript{2} reduction.}
    \label{fig:gen_trends}
\end{figure}

\noindent
Since the outcome of the decarbonization workflow is the baseline plus 13 systems,
each for a different level of decarbonization, it is more useful to visualize the
evolution for the key metrics, rather than go through them one by one.
In order to decarbonize the system completely the total installed generation
capacity increases from around 1400~GW at the baseline, to around 2500~GW in the
fully decarbonized system, or around a 78\% increase.
When the system decarbonizes the fossil fuel based electricity generation goes to
zero and is replaced primarily with offshore wind and solar PV generation,
as shown in Figure~\ref{fig:gen_trends}.
The generation dominance of off-shore wind at high levels of decarbonization is
clearly visible.
Interestingly at very high levels of decarbonization nuclear generation, as well as
wind and biomass declines.
The sudden decline in nuclear generation coincides with the large buildout of the
storage capacity and seems to indicate that the system transitions from being
dominated by one paradigm to being dominated by another paradigm.

\noindent
The contribution to the total transmission capacity from the HVDC links remains
fairly constant as the system decarbonizes, while the AC transmission capacity
increases substantially.
The total transmission capacity sees a large increase at very high levels of
decarbonization as the system relies more and more on non-dispatchable renewable
generation which must be transported from the generation sites to the load centers
or stored.
Above approximately 90\% decarbonisation, storage deployment accelerates sharply as dispatchable fossil-based generation is phased out and balancing increasingly depends on energy storage.

\begin{figure}[htbp]
    \centering
    \includegraphics[width=0.85\textwidth]{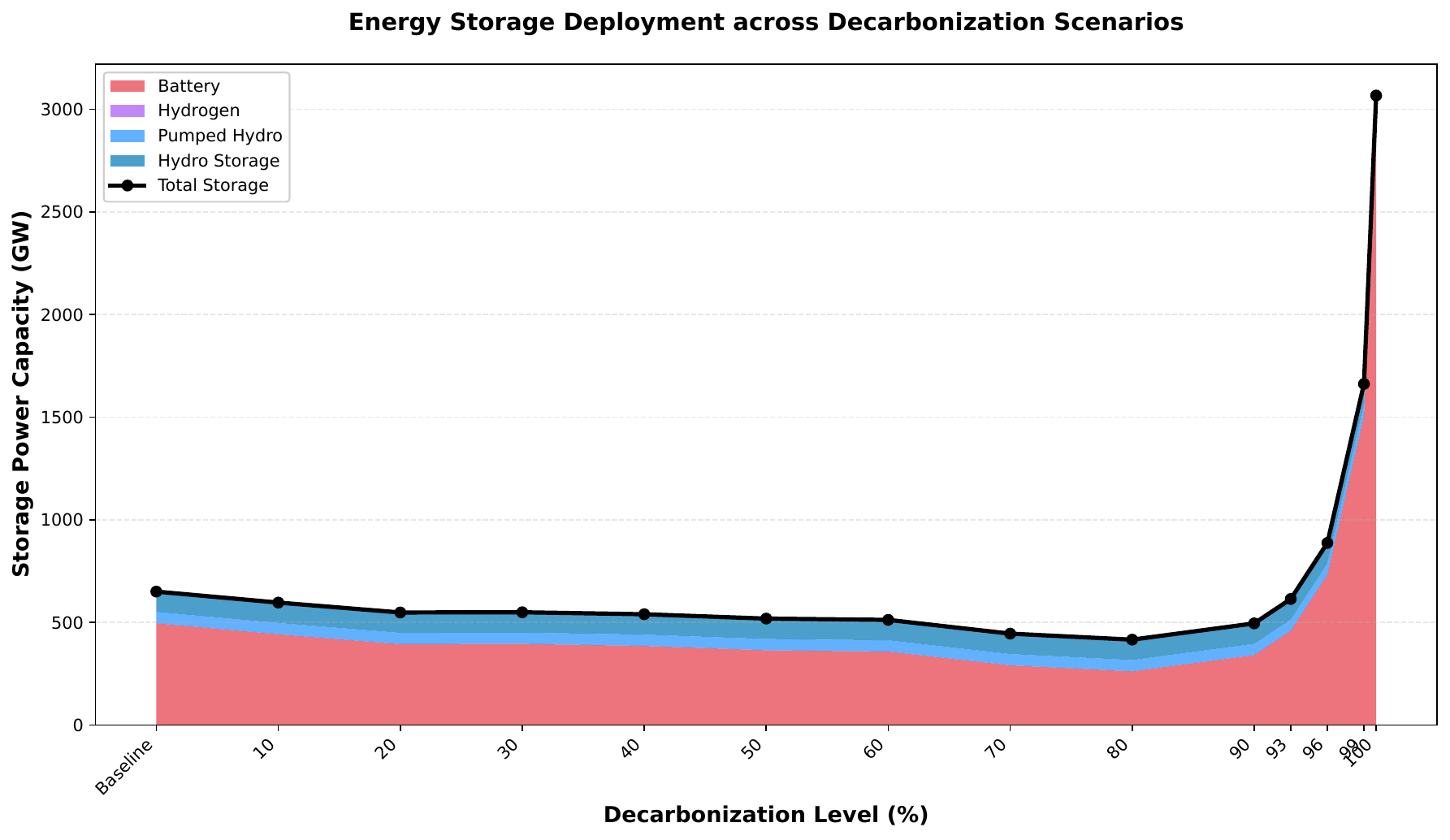}
    \caption{Energy storage deployment across decarbonisation scenarios, showing the sudden and massive increase in battery storage capacity above 90\% CO\textsubscript{2} reduction.}
    \label{fig:storage}
\end{figure}

\begin{figure}[htbp]
    \centering
    \includegraphics[width=0.85\textwidth]{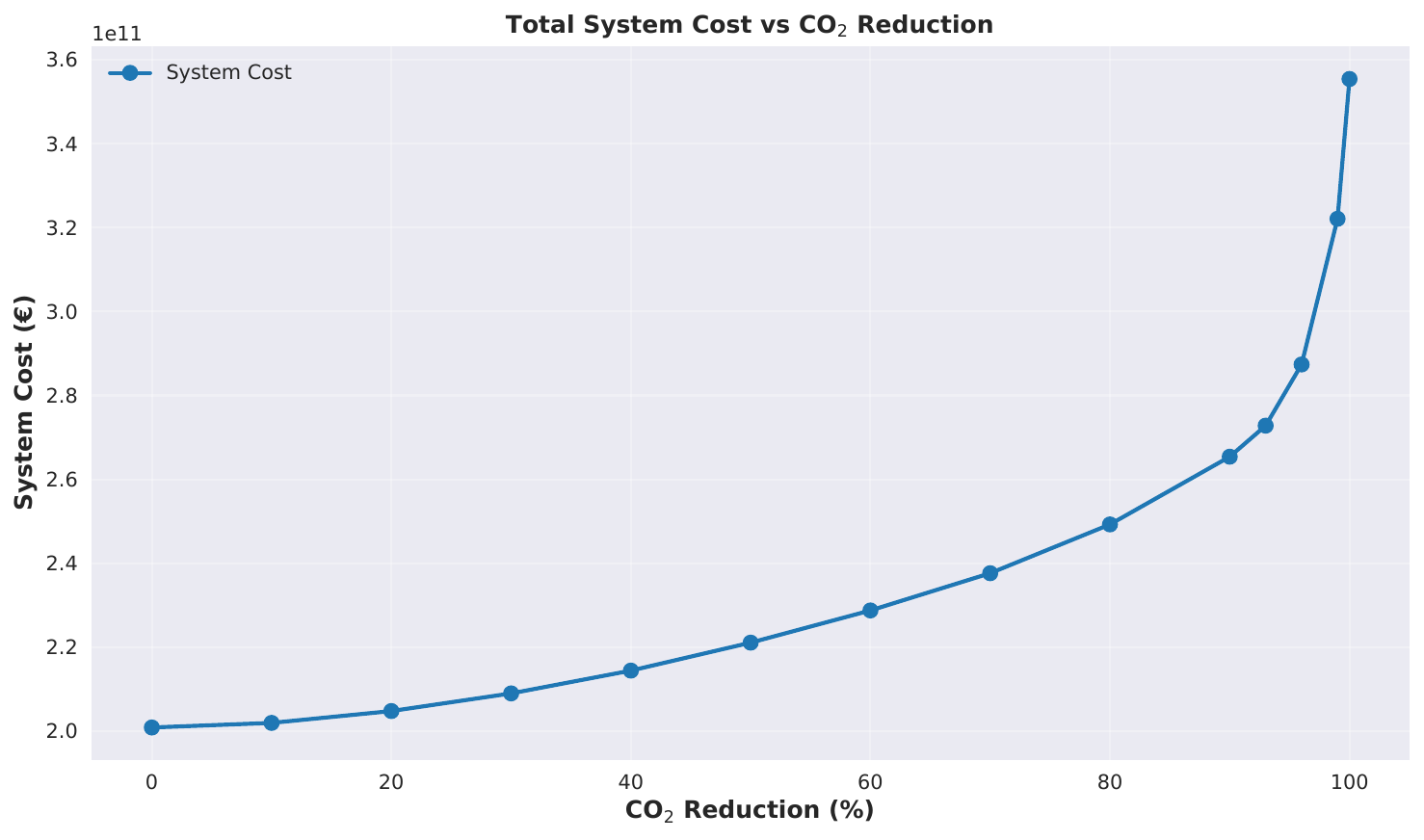}
    \caption{Total system cost vs.\ CO\textsubscript{2} reduction level, showing the non-linear increase in system cost with decarbonisation.}
    \label{fig:syscost}
\end{figure}

\begin{figure}[htbp]
    \centering
    \includegraphics[width=0.85\textwidth]{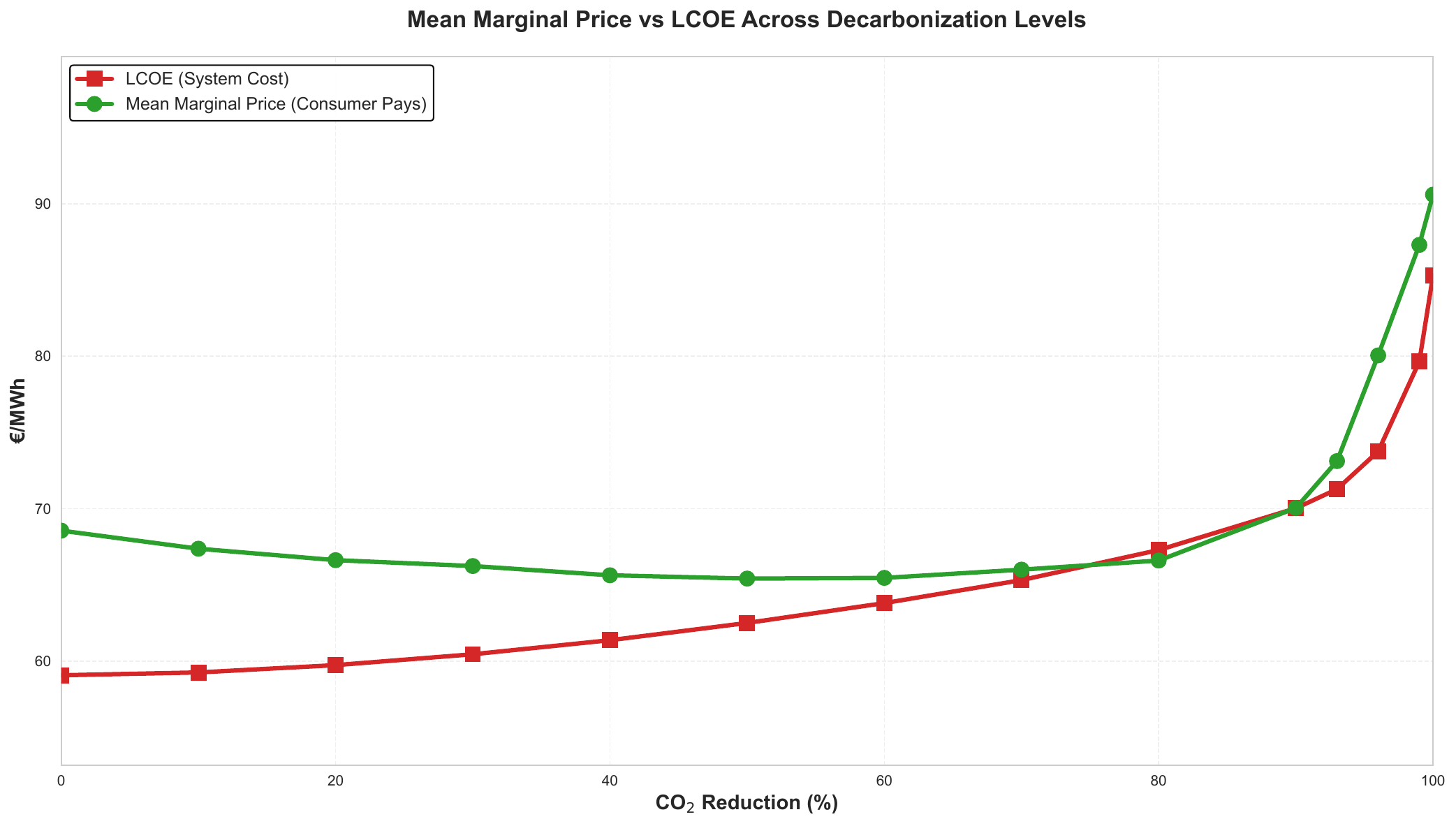}
    \caption{Comparison of mean LCOE and mean marginal demand-weighted price, illustrating scarcity rent evolution and the transition between dominant contributors at 80\% decarbonisation.}
    \label{fig:lcoe}
\end{figure}

\noindent
The mean marginal electricity price initially decreases during early decarbonisation due to merit-order effects associated with low-marginal-cost renewable generation. However, at high decarbonisation levels both marginal prices and total system costs increase rapidly as storage, transmission, and renewable overcapacity become dominant balancing mechanisms.
The total system cost increases quite dramatically from 200~billion~\euro\ at the
baseline to around 360~billion~\euro\ at 100\% decarbonization, corresponding to
a 80\% increase,
as seen in Figure~\ref{fig:syscost}.
The main contributor to this large increase is the large buildout of transmission-
and storage-infrastructure needed for balancing.
The divergence between the levelised cost of electricity and the mean marginal demand-weighted price, shown in Figure~\ref{fig:lcoe}, further illustrates how scarcity rents evolve as the system decarbonises.

\subsection{Transition in dominant optimisation behaviour at 80–90\% decarbonisation}

\noindent
The scarcity rents manifest when the optimizer ideally would increase capacity to
meet demand, but is constrained by a binding capacity constraint.
At low levels of decarbonization the majority of the contribution to the total
scarcity rent comes from the fossil-fuel based generators.
As the system decarbonizes the amount of fossil fuel generation is reduced, in
order to meet the decarbonization constraint, and as such the contribution to the
scarcity rent decreases.
However at around 80\% decarbonization, where the vast majority of the fossil
based generation has been phased out, the main contributors become the dispatchable
renewable generators, such as nuclear, and storage units needed for balancing.

\begin{figure}[htbp]
    \centering
    \includegraphics[width=0.85\textwidth]{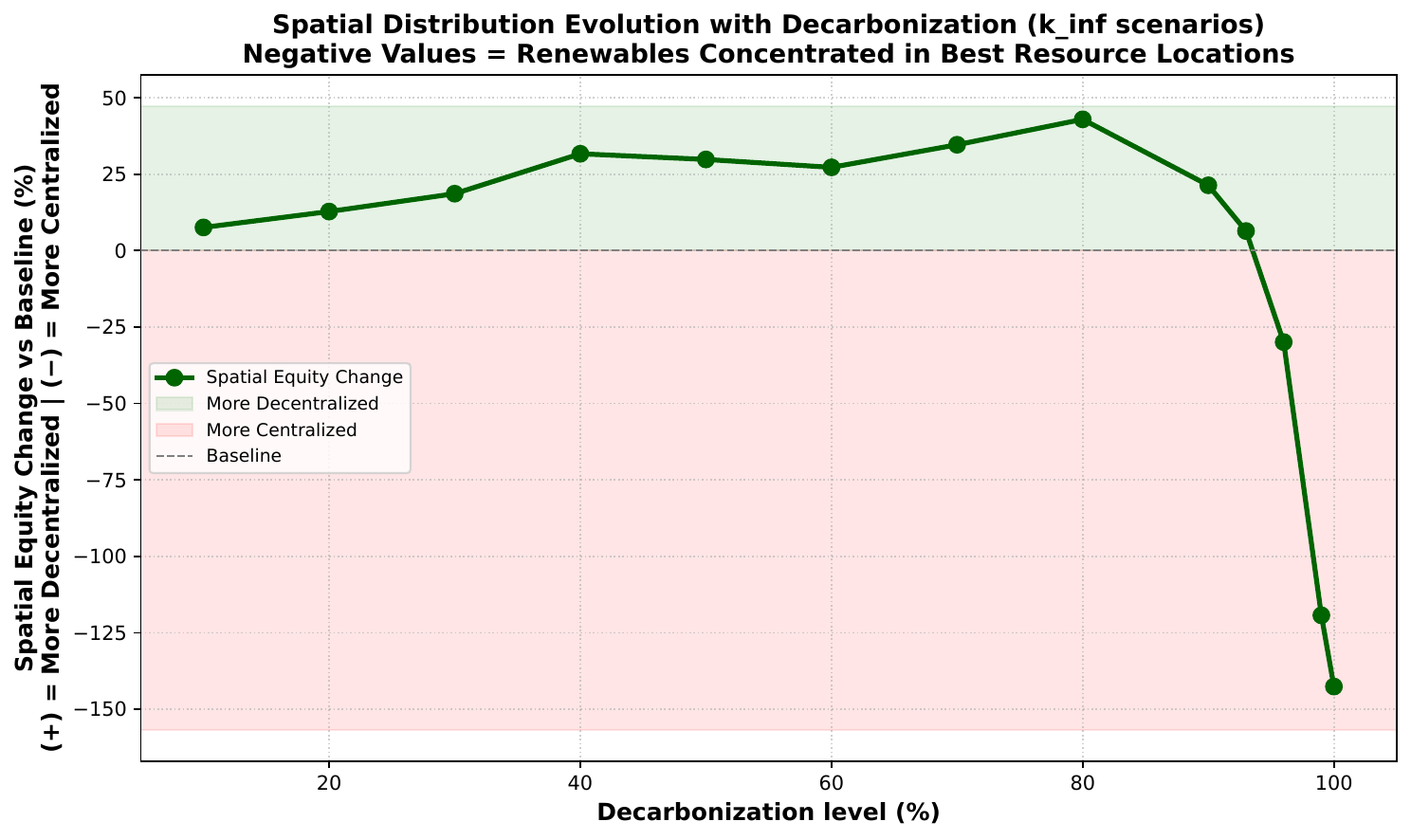}
    \caption{Spatial equity change relative to the baseline as a function of decarbonisation level, showing the initial improvement in equity followed by the sharp collapse above 80\% CO\textsubscript{2} reduction.}
    \label{fig:equity_evol}
\end{figure}

\begin{figure}[htbp]
    \centering
    \includegraphics[width=0.85\textwidth]{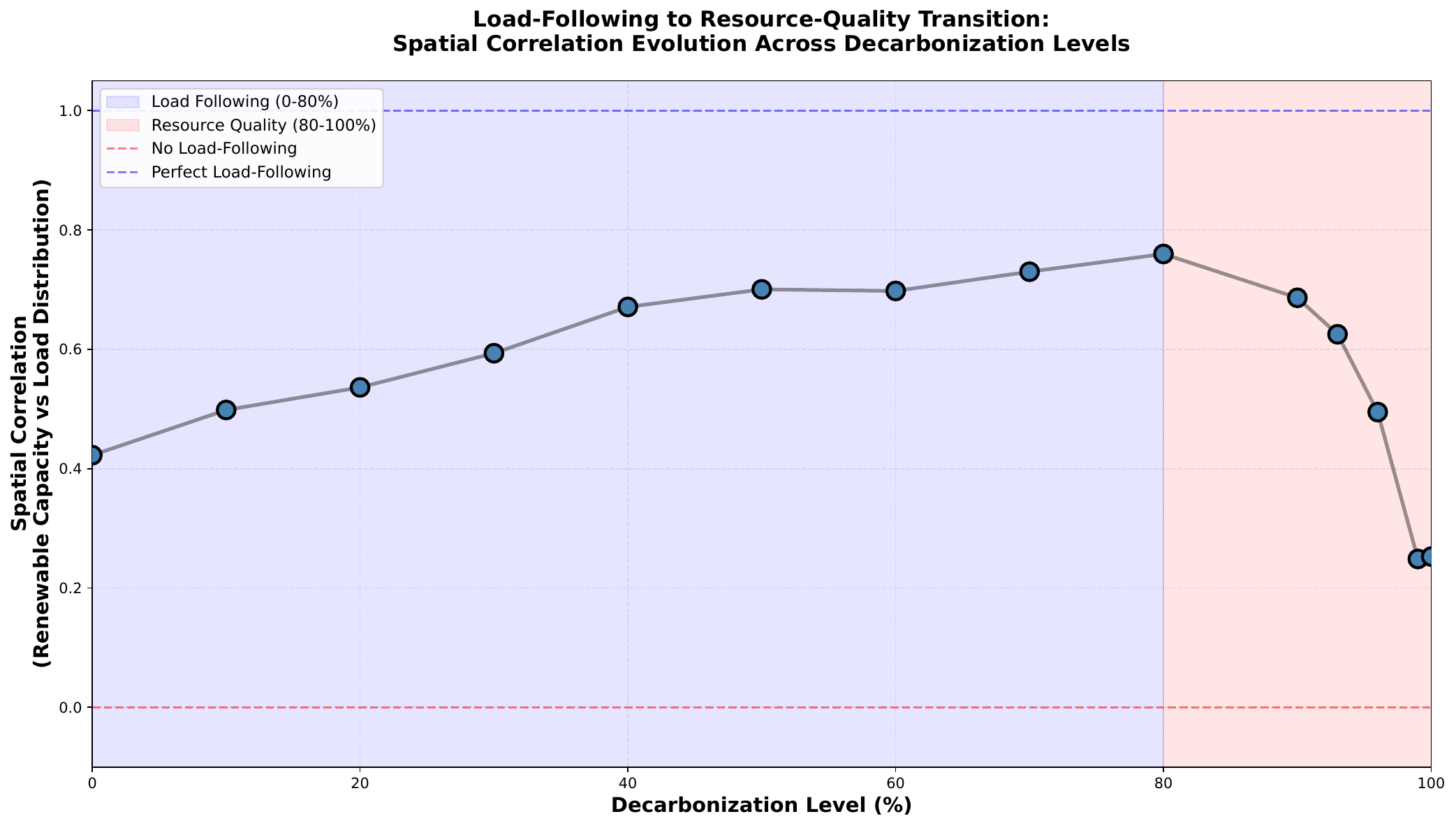}
    \caption{Spatial correlation between renewable capacity and load across decarbonisation levels, illustrating the transition from load-following (0--80\%) to resource-quality (80--100\%) behaviour.}
    \label{fig:spatial_corr}
\end{figure}

\begin{figure}[htbp]
    \centering
    \includegraphics[width=0.85\textwidth]{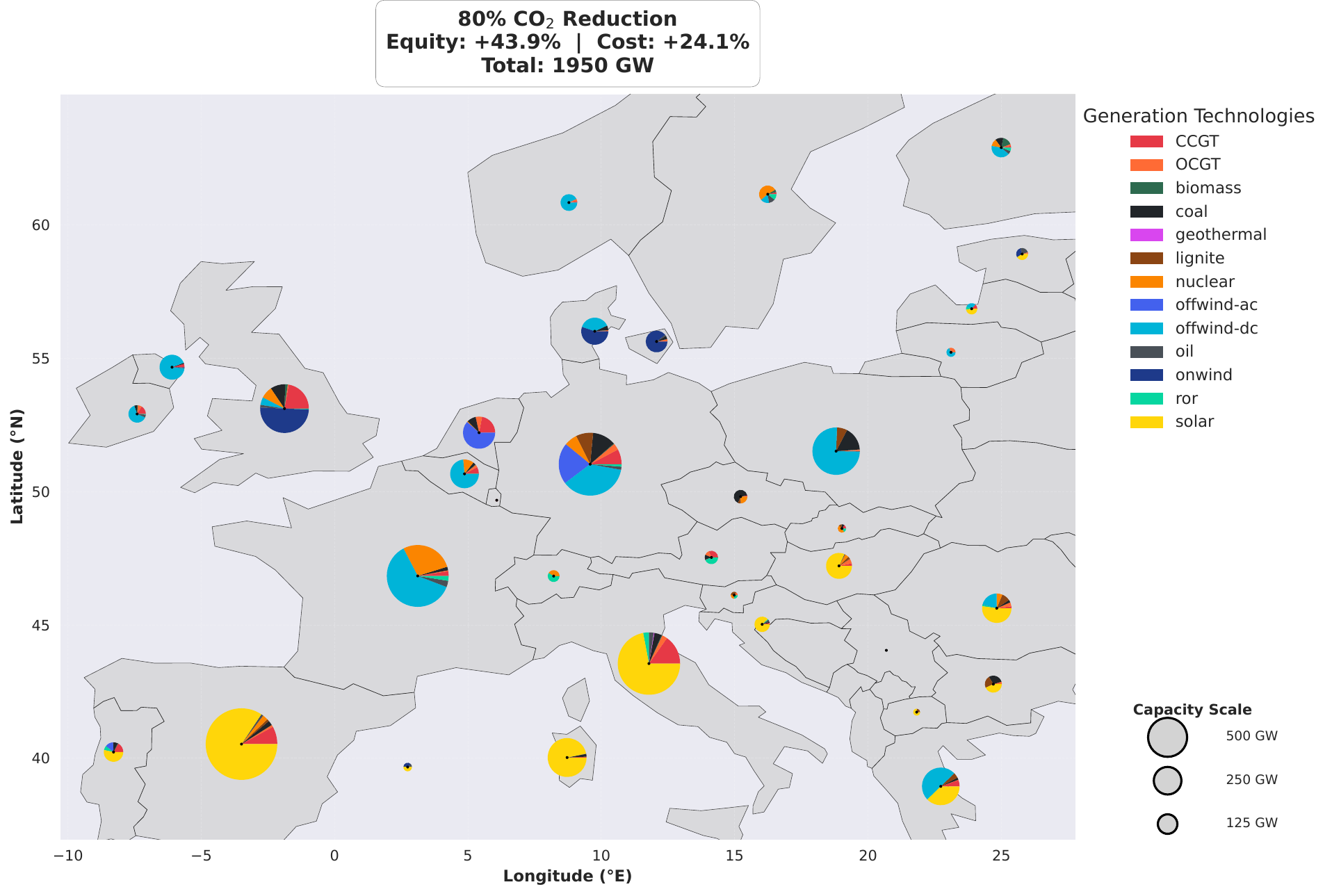}
    \caption{Generation capacity map at 80\% CO\textsubscript{2} reduction, showing the generation mix by node at the load-following to resource-quality transition point.}
    \label{fig:capmap80}
\end{figure}

\begin{figure}[htbp]
    \centering
    \includegraphics[width=0.85\textwidth]{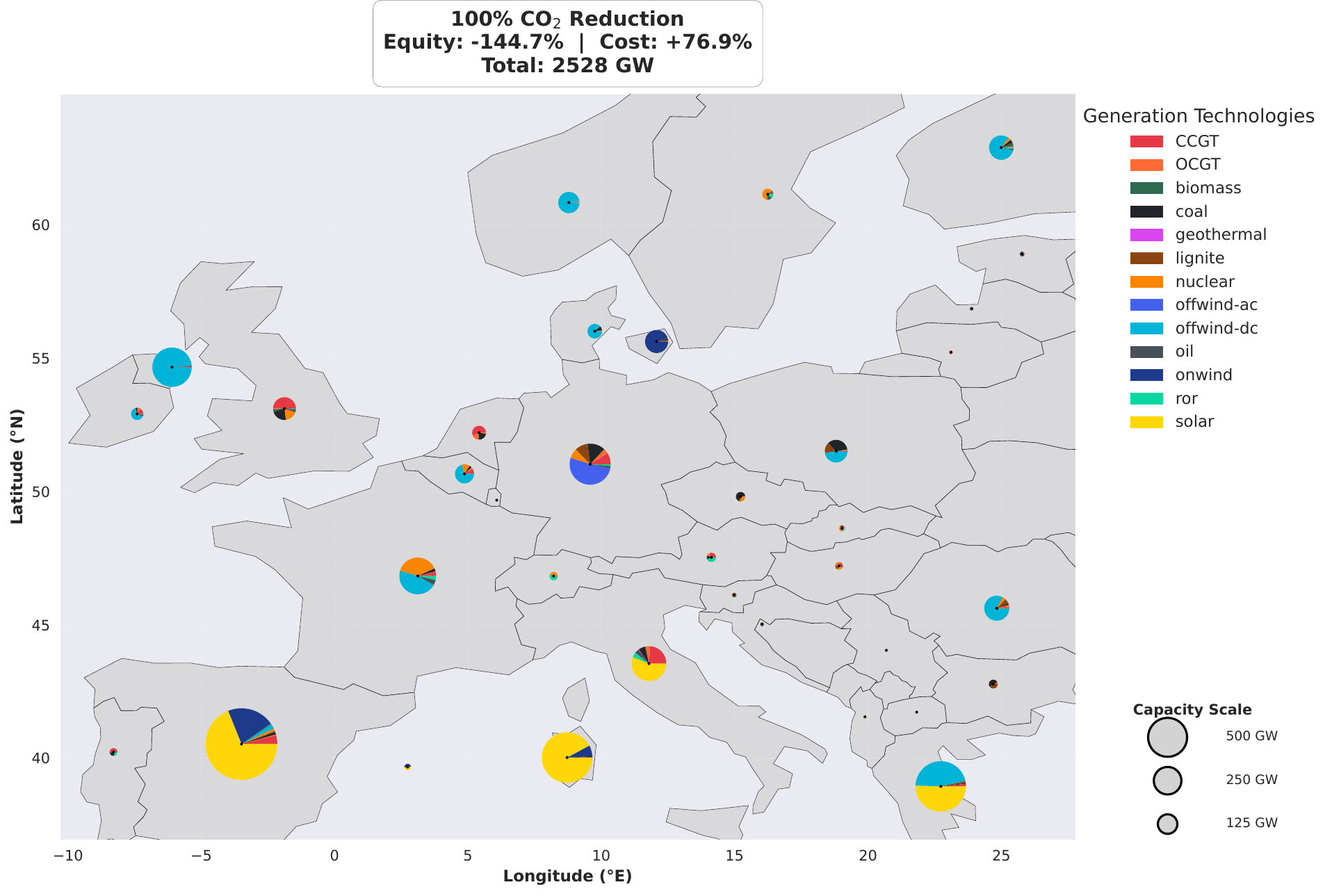}
    \caption{Generation capacity map at 100\% CO\textsubscript{2} reduction, showing the spatial concentration of renewable capacity in high-resource regions at full decarbonisation.}
    \label{fig:capmap100}
\end{figure}

As the system decarbonizes the equity initially increases, meaning that the system
becomes less centralized.
At lower levels of decarbonization the newly installed renewable generation
capacity is following the load centers in the network.
When compared to the baseline system this leads to more capacity being installed in
more diverse locations, and therefore increasing equity.
However at high levels of decarbonization the equity begins to collapse as the
system becomes highly centralized,
as shown in Figure~\ref{fig:equity_evol}.
At very high levels of decarbonization, where renewables dominate the system and
backup from dispatchable based generation decreases, the system begins to
increasingly optimize for the best renewable resources in order to minimize costs.
A pronounced transition from load-following renewable deployment toward increasingly resource-quality-driven planning emerges around  80--90\%
decarbonization where many of the metrics begin to change more rapidly.
This transition is further illustrated by the spatial correlation between installed renewable capacity and nodal load, shown in Figure~\ref{fig:spatial_corr}, and by the capacity maps at 80\% and 100\% decarbonisation in Figures~\ref{fig:capmap80} and~\ref{fig:capmap100}.
The transition region is additionally characterised by rapidly increasing transmission and storage requirements, reflecting the growing need to spatially and temporally balance highly renewable electricity generation.

\subsection{Decentralised systems --- technology mix and capacity factor degradation}

\begin{figure}[htbp]
    \centering
    \includegraphics[width=0.85\textwidth]{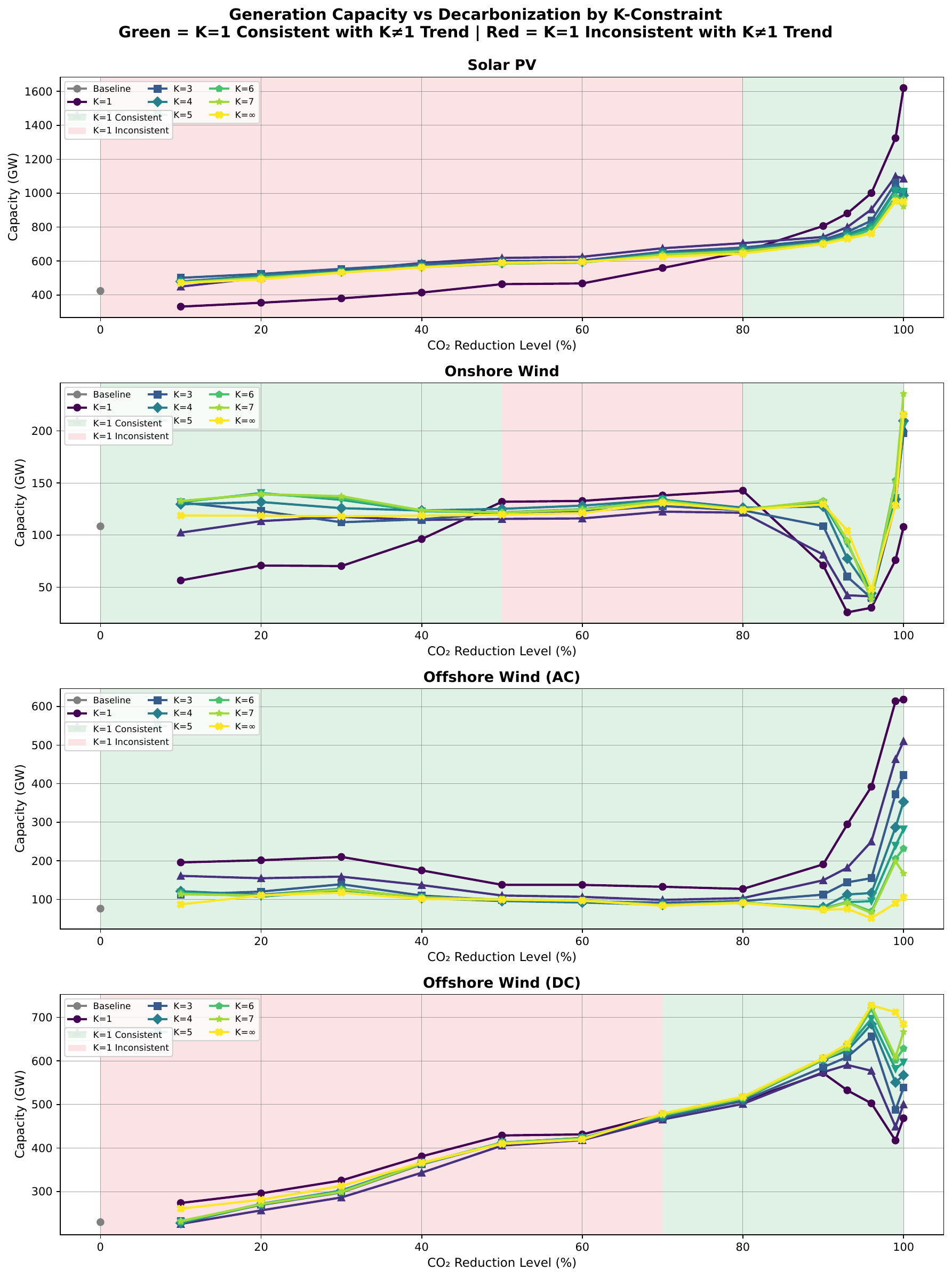}
    \caption{Total installed renewable generation capacity across K-constraints and decarbonisation levels, highlighting the divergence of the $K=1$ scenario from the general decentralisation trend.}
    \label{fig:ren_cap_k}
\end{figure}

\begin{figure}[htbp]
    \centering
    \includegraphics[width=0.85\textwidth]{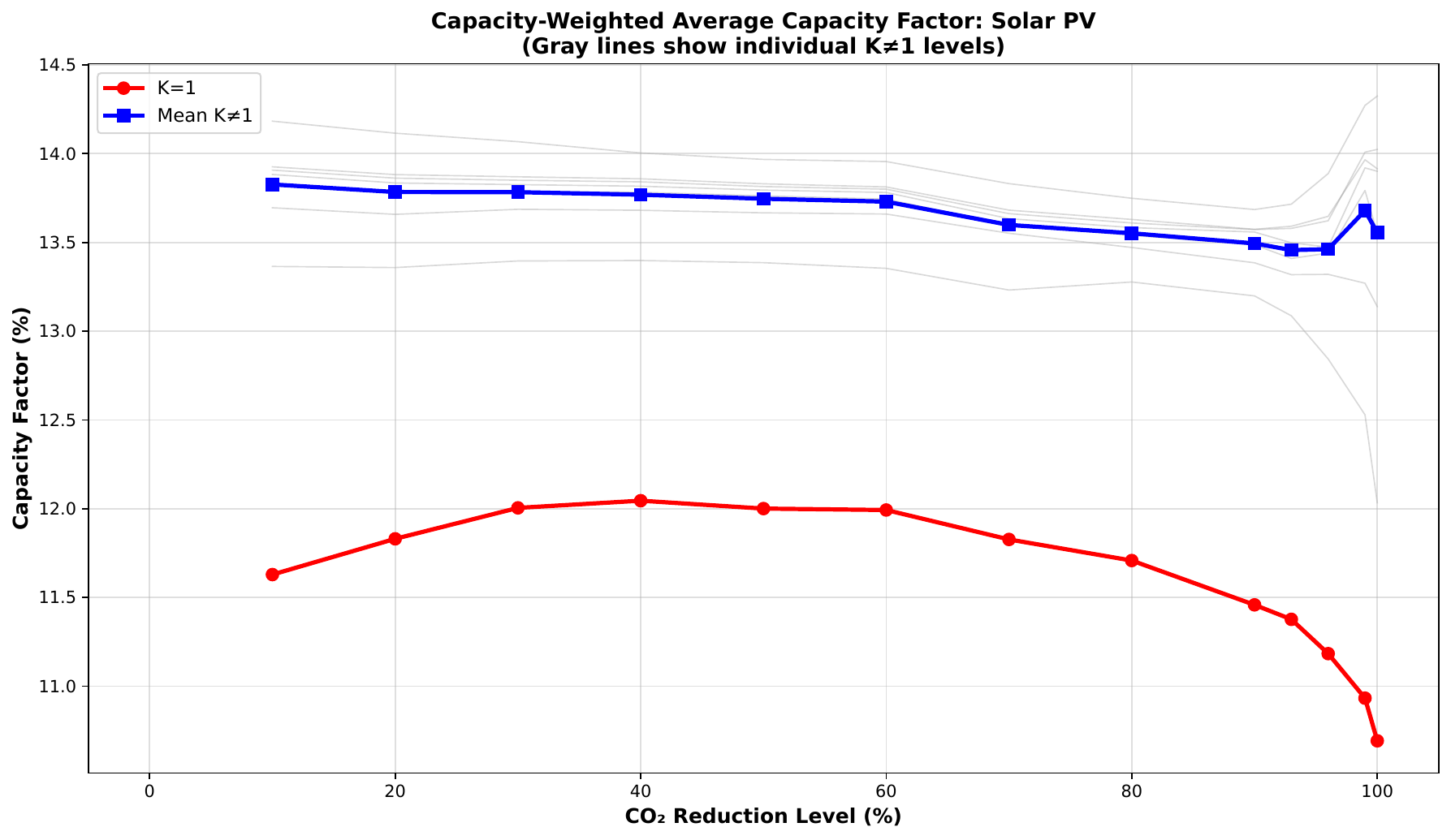}
    \caption{Capacity-weighted average capacity factor for solar PV across K-constraints and decarbonisation levels, showing the 13--21\% degradation under full decentralisation ($K=1$).}
    \label{fig:cf_solar}
\end{figure}

\noindent
Having performed perturbations of both the level of decarbonization and
decentralization constraint $K$, yields a total of 105 networks.
The generation capacity CV analysis, shows that the freely extendable renewable
generation technologies solar and wind are impacted the most.
Overall it seems that decentralization leads to an increase in solar and
AC-offshore wind capacity, as these can be more easily distributed across the
network, compared to onshore wind farms and DC-offshore wind,
as shown in Figure~\ref{fig:ren_cap_k}.
The mean capacity factor for solar remains fairly consistent at all levels of
decarbonization for $K \neq 1$.
However for the extreme $K=1$ scenario the mean capacity factor decreases
significantly as the system decarbonizes and is in general 13\%--21\% lower than
the mean of the other decentralization levels,
as shown in Figure~\ref{fig:cf_solar}.
Increasing decentralisation leads to deployment in progressively less favourable renewable locations, reducing average renewable capacity factors, particularly under the fully decentralised $K=1$ scenario.
The fully decentralised $K=1$ scenario behaves fundamentally differently from the remaining decentralisation levels, as the restrictive spatial allocation constraint increasingly overrides resource-quality optimisation.

\subsection{Cost-equity trade-off}

\begin{figure}[htbp]
    \centering
    \includegraphics[width=0.85\textwidth]{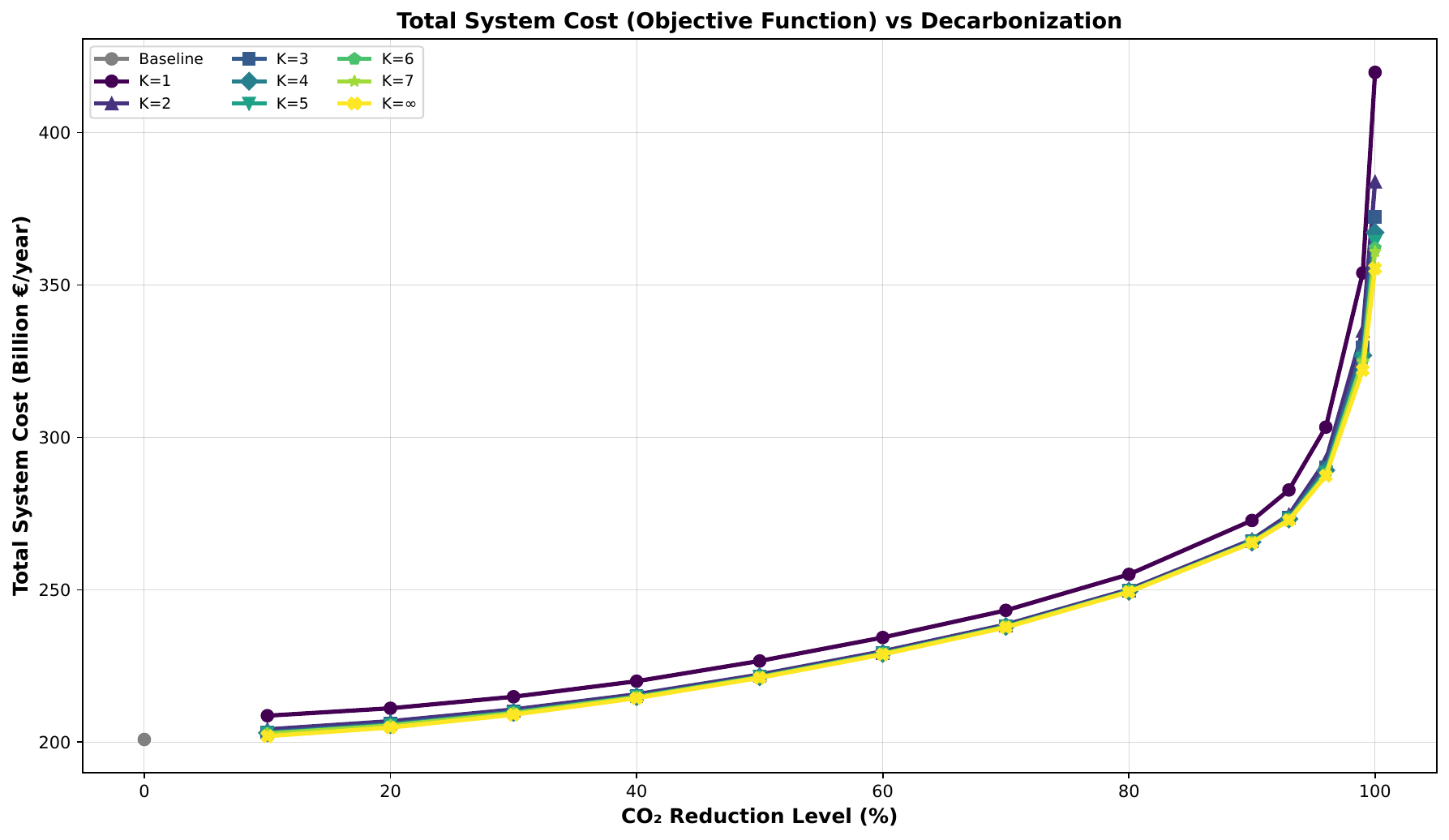}
    \caption{Total system cost across all K-constraints and decarbonisation levels.}
    \label{fig:syscost_k}
\end{figure}

\begin{figure}[htbp]
    \centering
    \includegraphics[width=0.85\textwidth]{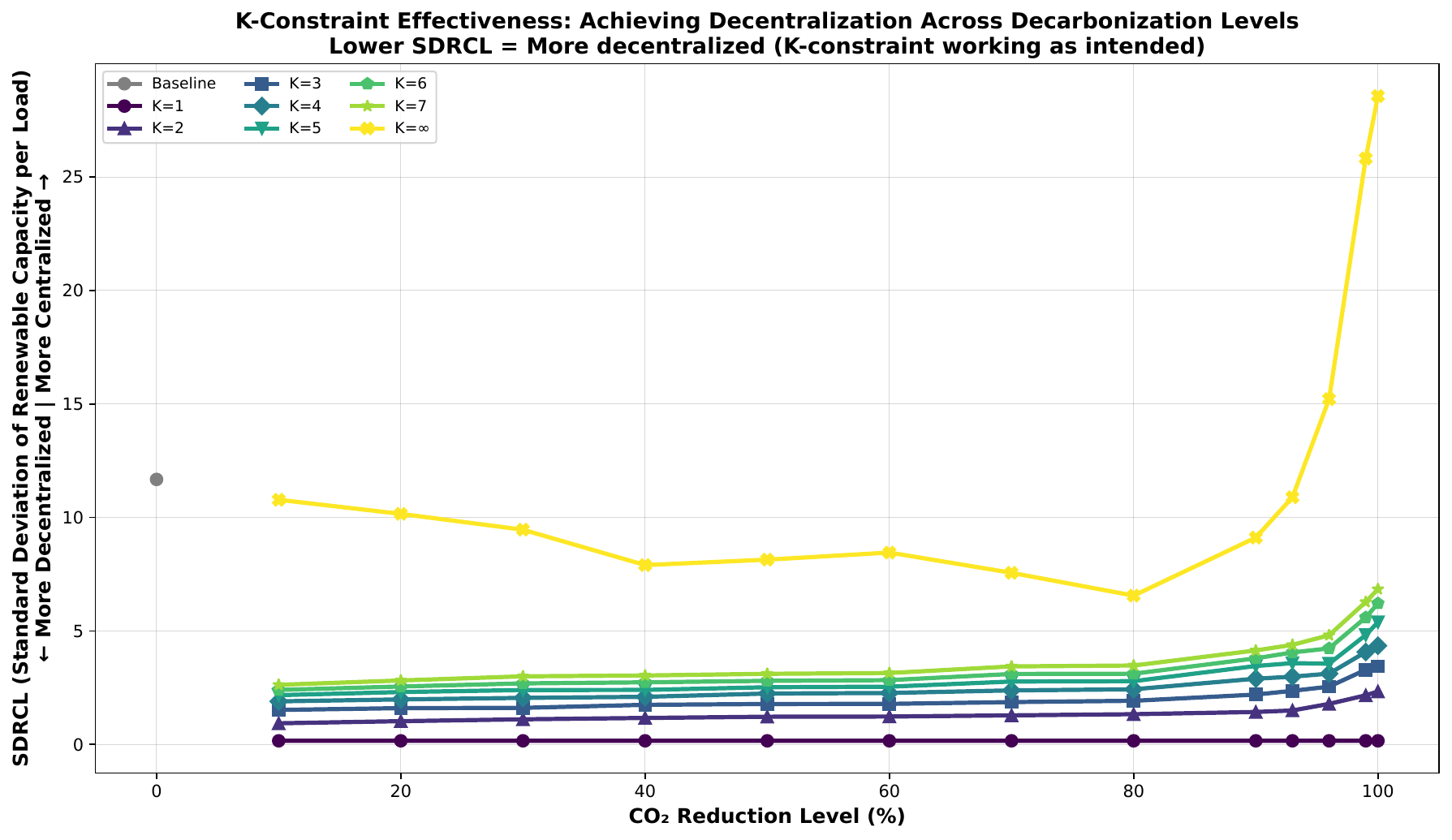}
    \caption{SDRCL equity measure across all K-constraints and decarbonisation levels, showing the flat cost solution space and the successful decentralisation across all scenarios.}
    \label{fig:equity_k}
\end{figure}

\begin{figure}[htbp]
    \centering
    \includegraphics[width=\textwidth]{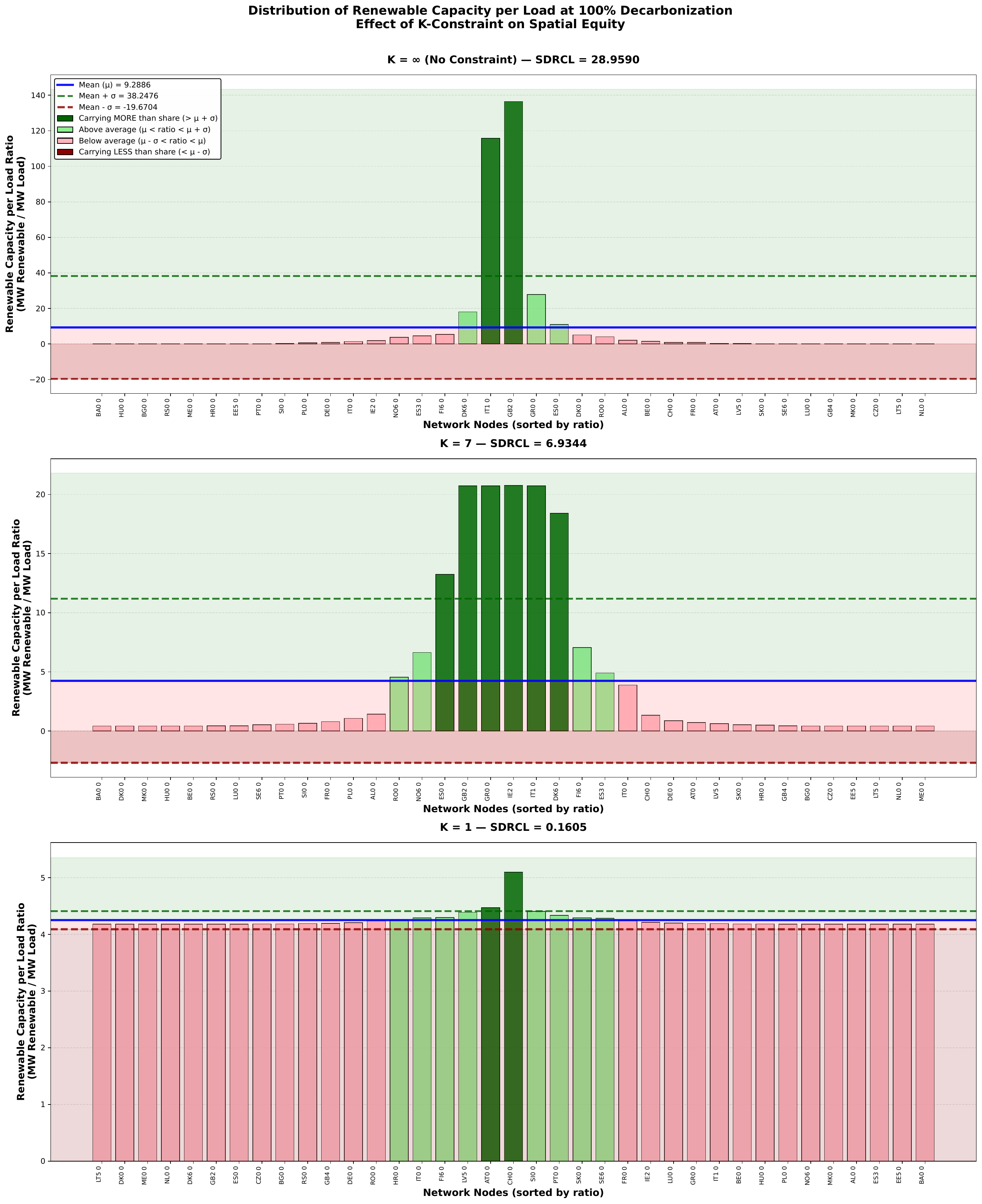}
    \caption{Distribution of renewable capacity to load ratio at 100\% decarbonisation for $K=\infty$, $K=7$, and $K=1$, illustrating how the K-constraint progressively reshapes the nodal capacity distribution towards spatial equity.}
    \label{fig:distribution}
\end{figure}

\noindent
Total system cost behaves in an intuitive manner, with an increase in
decentralization leading to a higher total system cost, especially at high levels
of decarbonization where the resource quality paradigm is in full effect,
as seen in Figure~\ref{fig:syscost_k}.
Forcing decentralization then means that the system can no longer fully utilize the
best renewable resources and must instead rely on less efficient and more expensive
local resources.
The K-constrained networks all achieve significant reduction in centralization,
while having similar cost premiums
as shown in Figure~\ref{fig:equity_k}.
The relatively small cost differences between intermediate K-values suggest a comparatively flat cost-equity trade-off region.
This suggests that moderate decentralisation can substantially improve spatial equity while imposing comparatively small additional system costs.
\medskip
\noindent
Critically, moderate decentralization (K=7) achieves 48--76\% of K=1's equity
benefits at only 2--12\% of K=1's cost penalty across the decarbonization levels,
demonstrating a 7--34$\times$ superior cost-equity trade-off.
This suggests partial decentralization is far more economically viable than
complete spatial equity.
Table~\ref{tab:tradeoff} summarises this cost-equity trade-off for selected decarbonisation levels, and Figure~\ref{fig:distribution} illustrates how the K-constraint progressively reshapes the nodal capacity distribution at 100\% decarbonisation.

\begin{table}[htbp]
\centering
\caption{Cost-equity trade-off for $K=7$ vs.\ $K=1$ vs.\ $K=\infty$ at selected decarbonisation levels. Cost premium and equity benefit are calculated relative to $K=\infty$. Efficiency is the ratio of equity benefit to cost premium.}
\label{tab:tradeoff}
\begin{tabular}{llrrrr}
\toprule
Decarb.\ level & K value & Cost premium (\%) & Equity benefit (\%) & Efficiency vs.\ $K=1$ & Efficiency vs.\ $K=\infty$ \\
\midrule
10\%  & $K=1$      & 3.32  & 98.5 & 1.0$\times$  & 29.7$\times$   \\
      & $K=7$      & 0.39  & 75.7 & 6.5$\times$  & 193.2$\times$  \\
\midrule
80\%  & $K=1$      & 2.32  & 97.6 & 1.0$\times$  & 42.1$\times$   \\
      & $K=7$      & 0.05  & 47.0 & 24.5$\times$ & 1029.3$\times$ \\
\midrule
100\% & $K=1$      & 18.11 & 99.4 & 1.0$\times$  & 5.5$\times$    \\
      & $K=7$      & 1.56  & 76.1 & 8.9$\times$  & 48.7$\times$   \\
\bottomrule
\end{tabular}
\end{table}

\section{Discussion and Policy Implications}

\noindent
The results demonstrate that least-cost deep decarbonisation pathways can conflict with spatial equity and resilience objectives, particularly at high renewable penetrations where system planning becomes increasingly resource-quality driven.

The analysis suggests that deep decarbonisation introduces a structural trade-off between least-cost optimisation and spatially balanced infrastructure deployment. While unconstrained optimisation minimises total system cost, it also increasingly concentrates renewable deployment in regions with the most favourable resources, leading to growing dependence on transmission infrastructure and cross-border balancing.

The transition observed around 80--90\% decarbonisation indicates a shift in the dominant optimisation behaviour governing the system. At lower decarbonisation levels, renewable deployment broadly follows major load centres, improving spatial equity relative to the baseline system. However, as dispatchable fossil-based generation is phased out, the optimisation increasingly prioritises regions with superior renewable resources in order to minimise balancing and infrastructure costs. This transition is associated with rapidly increasing storage and transmission requirements, as well as growing spatial concentration of renewable deployment.

The evolution of scarcity rents further reflects this transition in dominant system constraints. At lower decarbonisation levels, scarcity rents are primarily associated with fossil-based generation capacity, whereas at high decarbonisation levels they increasingly shift toward renewable generation, storage, and transmission infrastructure.

The results further demonstrate that moderate decentralisation constraints can substantially improve spatial equity without fundamentally disrupting least-cost decarbonisation pathways. In contrast, the fully decentralised K = 1 regime increasingly overrides resource-quality optimisation, leading to strong renewable overcapacity, declining capacity factors, and rapidly increasing balancing requirements.

The fully decentralised $K=1$ regime behaves fundamentally differently from the intermediate decentralisation levels because the restrictive spatial allocation constraint increasingly overrides resource-quality optimisation. This leads to declining renewable capacity factors, substantial renewable overcapacity, increased local balancing requirements, and significantly higher storage deployment. At high decarbonisation levels, additional solar deployment increasingly becomes the dominant balancing response under the fully decentralised regime due to the strong spatial correlation of wind generation across the network.

\medskip
\noindent
Considering these competing optimisation behaviours, a tradeoff between decentralization and cost
must be made.
It is however critical that this does not come at the expense of decarbonization
goals.
The results from this indicate that decentralization can be successfully enforced,
although at a cost premium ranging from 0.39\% at low levels of decarbonization
and decentralization to 18.11\% at high levels of decarbonization and
decentralization.
These results indicate that decentralisation and decarbonisation are not mutually exclusive objectives and can be achieved simultaneously, although at additional system cost.
Another sacrifice that must be made when enforcing decentralization is the
degradation of capacity factor for renewable generation technologies, as the
system is forced to install capacity in suboptimal locations.
This then leads to an increase in installed capacity and the cost premiums seen
above.
Finally decentralization also leads to the system becoming increasingly reliant on
energy storage in order to balance the system locally which again leads to
increased costs.

From a policy perspective, the results suggest that partially decentralised transition pathways may provide a compromise between economic efficiency, spatial equity, and infrastructure resilience in future European electricity systems.

The behaviour of the fully decentralised K = 1 regime suggests that extreme spatial equity enforcement may lead to increasingly inefficient and unstable planning outcomes under deep decarbonisation. Intermediate decentralisation levels therefore appear substantially more robust and economically viable than complete spatially uniform renewable deployment.

\subsection{Limitations and future work}

\noindent
Several limitations should be acknowledged. 
The model is limited to a single weather year and an electricity-only system without sector coupling, which may influence both balancing requirements and optimal spatial deployment patterns. Furthermore, the conservative hydrogen storage assumptions reduce the competitiveness of long-duration storage technologies within the model framework.

Future work could investigate alternative equity metrics, sector-coupled systems, multiple weather years, and higher spatial resolutions in order to evaluate the robustness of the identified planning regimes and cost-equity trade-offs.

\section{Conclusion}

\noindent
Deep decarbonisation substantially reshapes the structure of the European electricity system, increasing reliance on renewable generation, storage, and transmission infrastructure. A pronounced transition was identified around 80--90\% decarbonisation, where the optimisation shifts from load-following renewable deployment toward increasingly resource-quality-driven planning.

Without decentralisation constraints, this transition leads to strong spatial concentration of renewable deployment in regions with favourable renewable resources. While economically efficient, such pathways may increase infrastructure dependence and reduce spatial equity across the system.

The proposed K-constraint framework demonstrates that moderate decentralisation can substantially improve spatial equity at comparatively small additional system cost. In contrast, fully decentralised systems increasingly override resource-quality optimisation, resulting in declining renewable capacity factors and rapidly increasing balancing requirements.

Overall, the results suggest that partially decentralised transition pathways may represent a viable compromise between least-cost optimisation, spatial equity, and infrastructure resilience in future European electricity systems.


\bibliographystyle{elsarticle-num}
\bibliography{Sources}

\appendix
\section{Technology cost assumptions}

Detailed techno-economic assumptions used in the optimisation model are shown in Tables \ref{tab:gencosts} and \ref{tab:storagecosts}.

\begin{table}[htbp]
\centering
\caption{Cost assumptions for generation technologies \cite{dea2023,iea2023}. Lignite is proxied with coal CAPEX. FOM reported as a rate is applied as a fraction of CAPEX per annum. 
}
\label{tab:gencosts}
\begin{tabular}{lrrrr}
\toprule
Carrier & CAPEX (EUR/MW) & Lifetime (yr) & Discount rate & FOM \\
\midrule
CCGT       & 830,000   & 25 & 0.07 & 27,800 EUR/MW/a \\
OCGT       & 435,000   & 25 & 0.07 & 7,745 EUR/MW/a  \\
Coal       & 1,860,000 & 30 & 0.07 & 30,355 EUR/MW/a \\
Lignite    & 1,860,000 & 30 & 0.07 & 30,355 EUR/MW/a \\
Oil        & 343,000   & 25 & 0.07 & 8,448 EUR/MW/a  \\
Biomass    & 3,300,000 & 25 & 0.07 & 96,000 EUR/MW/a \\
Geothermal & 4,290,000 & 20 & 0.07 & 2\% rate        \\
Nuclear    & 5,870,000 & 60 & 0.07 & 2\% rate        \\
\bottomrule
\end{tabular}
\end{table}

\begin{table}[htbp]
\centering
\caption{Storage technology cost assumptions \cite{dea2023}.}
\label{tab:storagecosts}
\begin{tabular}{llrr}
\toprule
Technology & Component & CAPEX (EUR/unit) & Annualised cost \\
\midrule
Battery   & Energy (cells)       & 142/kWh  & 12.2k EUR/MWh/yr \\
          & Power (inverter)     & 160/kW   & 13.7k EUR/MW/yr  \\
Hydrogen  & Energy (tank)        & 57/kWh   & 4.9k EUR/MWh/yr  \\
          & Power (electrolyser) & 600/kW   & 48.4k EUR/MW/yr  \\
          & Power (fuel cell)    & 1100/kW  & 156.6k EUR/MW/yr \\
\bottomrule
\end{tabular}
\end{table}

\end{document}